\documentstyle{article}
\begin{document}

\title{Quenched Disorder From Sea-Bosons}
\author{ Girish S. Setlur \\ The Institute of Mathematical Sciences \\
 Taramani, Chennai 600113 }
\maketitle                 

\begin{abstract}
 The degenerate Fermi gas coupled to a random potential is used
 to study metal-insulator transitions in various dimensions.
 We first recast the problem in the sea-boson language that
 allows for an easy evaluation of important physical attributes. 
 We evaluate the dynamical number-number correlation function
 and from this compute the a.c. conductivity. 
 We find that the d.c.
 conductivity vanishes in one and two dimensions.
 For a hamiltonian that forbids scattering of an electron from within 
 the Fermi surface to another state within the Fermi surface we find
 that there is no metal-insulator transition in three dimensions either.  
\end{abstract} 

\section{Introduction}

   In a series of published works \cite{Setlur1} \cite{Setlur2}, 
   and a recent preprint\cite{Setlur3},
   we showed how to extract the anomalous exponents
   in case of the Luttinger model using sea-bosons. This paves the way 
   for application of the amended sea-boson theory that now is powerful
   enough to reproduce most of the exactly known results in 1d, to 
   other systems such as electrons with quenched disorder
   with and without Coulomb interactions in various dimensions.
   The relevant literature on this subject is vast 
   and we shall not attempt to be exhaustive in surveying it.
   Anderson's pioneering
   work on localization\cite{Anderson} was followed by the 
   work of Abrahams et.al. \cite{Abra} and later on a more rigorous formulation
   of the notion of disorder averaging was given by McKane and
   Stone\cite{McKane}.
   This relates to a single electron in a disordered potential. 
   The classic review of Lee and Ramakrishnan \cite{Lee}
   includes many references on the
   literature concerning the degenerate electron gas in a disordered 
   potential. A more recent review is by Abrahams et. al. \cite{Abra}.

\section{Number-Number Correlation Function}

 Eventually, we would like to compute the a.c.
 conductivity at absolute zero. Unfortunately this quantity is rather
 difficult to compute. This is because it involves first calculating the  
 dynamical number-number correlation function. This latter function has proved
 very difficult to evaluate. Before we evaluate this quantity we would
 like to say a few words about how the dynamical number-number correlation
 enters into the picture. It is defined as follows.
\begin{equation}
N({\bf{k}}t;{\bf{k}}^{'}t^{'}) = <  n_{ {\bf{k}} }(t) 
n_{ {\bf{k}}^{'} }(t^{'}) >
 - < n_{ {\bf{k}} }(t) >
< n_{ {\bf{k}}^{'} }(t^{'})  >
\end{equation}
 Notice that the a.c. conductivity is related to the 
 dynamical total momentum-momentum correlation function. 
 This formula was derived in an earlier preprint\cite{Setlur3}. 
 The momentum-momentum correlation function in turn may be related
 to the dynamical number-number correlation function.
\begin{equation}
\left<  \delta {\hat{ {\bf{P}} }}(t)
\cdot \mbox{      }\delta {\hat{ {\bf{P}} }}(0) \right>
 = \sum_{ {\bf{k}} {\bf{k}}^{'} }
({\bf{k}} \cdot {\bf{k}}^{'}) 
 \mbox{    }
\left[ \left< n_{ {\bf{k}} }(t) \mbox{    }n_{ {\bf{k}}^{'} }(0) \right>
 - \left< n_{ {\bf{k}} }(t) \right>
 \left< n_{ {\bf{k}}^{'} }(0) \right> \right]
\end{equation}
 In a recent preprint\cite{Setlur3}, 
 we provided some hints as to how might go about computing the number-number
 correlation function for the interacting system. 
 It involves functional differentiation of the average momentum distribution
 with respect to sources that couple to the number operator. When this is done
 carefully we find the following appealing
 form of the dynamical number-number correlation function.
\[
N({\bf{k}}t;{\bf{k}}^{'}t^{'}) = 
 (1-n_{F}({\bf{k}}))(1-n_{F}({\bf{k}}^{'}))
S_{AA}({\bf{k}}t;{\bf{k}}^{'}t^{'})
 + n_{F}({\bf{k}}) n_{F}({\bf{k}}^{'})
S_{BB}({\bf{k}}t;{\bf{k}}^{'}t^{'})
\]
\begin{equation}
- (1- n_{F}({\bf{k}}))
 n_{F}({\bf{k}}^{'})                             
S_{AB}({\bf{k}}t;{\bf{k}}^{'}t^{'})
- (1- n_{F}({\bf{k}}^{'})) n_{F}({\bf{k}})                     
S_{BA}({\bf{k}}t;{\bf{k}}^{'}t^{'})
\end{equation}
Here the various quantities are defined recursively. First (here
 $ m,n = A,B $),
\begin{equation}
S_{mn}({\bf{k}}t;{\bf{k}}^{'}t^{'}) = 
e^{-2< {\hat{S}}_{m}({\bf{k}}) > }
\mbox{    }e^{-2< {\hat{S}}_{n}({\bf{k}}^{'}) > }
\mbox{     }S^{0}_{mn}({\bf{k}}t;{\bf{k}}^{'}t^{'})
\end{equation}
 One could take the point of
 view that $ S^{0}_{mn}({\bf{k}}t;{\bf{k}}^{'}t^{'}) $
 is evaluated by assuming that $ a_{ {\bf{k}} }({\bf{q}}) $ are canonical
 bosons, dropping all the square roots and so on.
 The reason being that the corrections caused by fluctuations in
 the momentum distributions are included in the exponential prefactors.
 These quantities are defined recursively.
\begin{equation}
{\hat{S}}_{A}({\bf{k}},t) = \sum_{ {\bf{q}} }
 A^{\dagger}_{ {\bf{k}}-{\bf{q}}/2 }({\bf{q}},t) 
A_{ {\bf{k}}-{\bf{q}}/2 }({\bf{q}},t) 
\end{equation}
\begin{equation}
{\hat{S}}_{B}({\bf{k}},t) = \sum_{ {\bf{q}} }
 A^{\dagger}_{ {\bf{k}}+{\bf{q}}/2 }({\bf{q}},t) 
A_{ {\bf{k}}+{\bf{q}}/2 }({\bf{q}},t) 
\end{equation}
\begin{equation}
S^{0}_{mn}({\bf{k}}t;{\bf{k}}^{'}t^{'})
 = \left< {\hat{S}}_{m}({\bf{k}},t) {\hat{S}}_{n}({\bf{k}}^{'},t^{'}) \right>
 -  \left< {\hat{S}}_{m}({\bf{k}},t) \right> 
 \left< {\hat{S}}_{n}({\bf{k}}^{'},t^{'}) \right>
\end{equation}
where $ m,n = A, B $.

\section{ The Toy Hamiltonian }

 Here we couple the free Fermi gas to a disorder potential and compute the a.c.
 conductivity. The diagonalization is
 rendered trivial in the sea-boson language.
 However, the formula for the dynamical number-number correlation function
 in terms of the bosons is very nontrivial
 and can therefore be expected to lead to nontrivial results.
\begin{equation}
H = \sum_{ {\bf{k}} {\bf{q}} } \frac{ {\bf{k.q}} }{m}
 A^{\dagger}_{ {\bf{k}} }({\bf{q}})A_{ {\bf{k}} }({\bf{q}})
 + \sum_{ {\bf{q}} }\frac{ U_{dis}({\bf{q}}) }{ \sqrt{V} }
 \sum_{ {\bf{k}} }\left[ A_{ {\bf{k}} }(-{\bf{q}}) + A^{\dagger}_{ {\bf{k}} }({\bf{q}}) \right]
\label{BOSETOY}
\end{equation}
 The above hamiltonian describes electrons close to the
 Fermi surface interacting with the disorder potential. However, notice
 that no externally chosen cutoff is needed. A natural smooth cutoff
 emerges by not linearizing the bare fermion dispersion. 
 In the Fermi language, Eq.(~\ref{BOSETOY}) is equivalent
 to the following hamiltonian.
\begin{equation}
H = \sum_{ {\bf{k}} } \epsilon_{ {\bf{k}} } c^{\dagger}_{ {\bf{k}} }
c_{ {\bf{k}} }
 + \sum_{ {\bf{k}}, {\bf{q}} }\frac{ U_{dis}({\bf{q}}) }{ \sqrt{V} }
\left[ \Lambda_{ {\bf{k}} }(-{\bf{q}}) + \Lambda_{ {\bf{k}} }({\bf{q}}) \right]
c^{\dagger}_{ {\bf{k}} + {\bf{q}}/2 } c_{ {\bf{k}} - {\bf{q}}/2 }
\label{TOYHAMIL}
\end{equation}
Here $ \Lambda_{ {\bf{k}} }({\bf{q}}) = 
n_{F}({\bf{k}}+{\bf{q}}/2)(1-n_{F}({\bf{k}}-{\bf{q}}/2)) $
 and $ n_{F}({\bf{k}}) = \theta(k_{F}-|{\bf{k}}|) $.
 Thus the toy hamiltonian Eq.(~\ref{TOYHAMIL}) describes electrons
 coupling to the disorder potential near the Fermi
 surface in such a way that processes
 that take an electron below the Fermi surface and place it in another
 state also below the Fermi surface or both above the Fermi surface
 are forbidden. We shall see that in this case there is no metal
 insulator transition in any dimension. However, we reproduce the
 results that in one and two dimensions, the d.c. conductivity is zero.  
 This hamiltonian may be trivially diagonalized by the
 following transformation.
\begin{equation}
A_{ {\bf{k}} }({\bf{q}},t) = A^{0}_{ {\bf{k}} }({\bf{q}})
e^{ -i \frac{ {\bf{k.q}} }{m} t } 
 - \frac{ U_{dis}({\bf{q}}) }{ \sqrt{V} } \frac{ m }{ {\bf{k.q}} }
\end{equation}
Thus we have,
\begin{equation}
\left< A^{\dagger}_{ {\bf{k}} }({\bf{q}}) A_{ {\bf{k}} }({\bf{q}}) \right>
 =  \frac{ |U_{dis}({\bf{q}})|^2 }{ V } \frac{ m^2 }{ ({\bf{k.q}})^2 }
n_{F}({\bf{k}}-{\bf{q}}/2)(1-n_{F}({\bf{k}}+{\bf{q}}/2))
\end{equation}
 Also for the number fluctuations,
\[
\left< A^{\dagger}_{ {\bf{k}} }({\bf{q}},t) A_{ {\bf{k}} }({\bf{q}},t) 
A^{\dagger}_{ {\bf{k}}^{'} }({\bf{q}}^{'},t^{'}) A_{ {\bf{k}}^{'} }
({\bf{q}}^{'},t^{'})  \right>
 - \left< A^{\dagger}_{ {\bf{k}} }({\bf{q}},t) A_{ {\bf{k}} }({\bf{q}},t) 
\right>
 \left< A^{\dagger}_{ {\bf{k}}^{'} }({\bf{q}}^{'},t^{'}) A_{ {\bf{k}}^{'} }
({\bf{q}}^{'},t^{'})  \right>
\]
\begin{equation}
 = \frac{ |U_{dis}({\bf{q}})|^2 }{ V } \frac{ m^2 }{ ({\bf{k.q}})^2 }
\mbox{     }
e^{ -i \frac{ {\bf{k.q}} }{m} (t-t^{'}) }
\delta_{ {\bf{k}},{\bf{k}}^{'} }\delta_{ {\bf{q}}, {\bf{q}}^{'} }
n_{F}({\bf{k}}-{\bf{q}}/2)(1-n_{F}({\bf{k}}+{\bf{q}}/2))  
\end{equation}
Thus we may compute the following quantities,
\begin{equation}
< {\hat{S}}_{A}({\bf{k}}) > = \sum_{ {\bf{q}} }
 \frac{ |U_{dis}({\bf{q}})|^2 }{ V }
 \frac{ m^2 }{ ( ({\bf{k}}-{\bf{q}}/2).{\bf{q}} )^2 }
 n_{F}({\bf{k}}-{\bf{q}})(1-n_{F}({\bf{k}}))
\end{equation}
\begin{equation}
< {\hat{S}}_{B}({\bf{k}}) > = \sum_{ {\bf{q}} }
 \frac{ |U_{dis}({\bf{q}})|^2 }{ V }
 \frac{ m^2 }{ ( ({\bf{k}}+{\bf{q}}/2).{\bf{q}} )^2 }
 n_{F}({\bf{k}})(1-n_{F}( {\bf{k}} + {\bf{q}} ))
\end{equation}
\begin{equation}
S_{AA}({\bf{k}},t ; {\bf{k}}^{'},t^{'}) 
 = \delta_{ {\bf{k}}, {\bf{k}}^{'} } \mbox{       }\sum_{ {\bf{q}} }
\frac{ |U_{dis}({\bf{q}})|^2 }{ V }
 \frac{ 1 }{ ( \frac{ {\bf{k.q}} }{m} - \epsilon_{ {\bf{q}} } )^2  }
\mbox{     }
e^{ -i ( \frac{ {\bf{k.q}} }{m} - \epsilon_{ {\bf{q}} } )(t-t^{'}) }
\mbox{     }
n_{F}({\bf{k}}-{\bf{q}})(1-n_{F}({\bf{k}}))  
\end{equation}
\begin{equation}
S_{BB}({\bf{k}},t ; {\bf{k}}^{'},t^{'}) 
 = \delta_{ {\bf{k}}, {\bf{k}}^{'} } \mbox{       } \sum_{ {\bf{q}} }
\frac{ |U_{dis}({\bf{q}})|^2 }{ V }
 \frac{ 1 }{ ( \frac{ {\bf{k.q}} }{m} + \epsilon_{ {\bf{q}} } )^2  }
\mbox{     }
e^{ -i ( \frac{ {\bf{k.q}} }{m} + \epsilon_{ {\bf{q}} } )(t-t^{'}) }
\mbox{        }
n_{F}({\bf{k}})( 1 - n_{F}({\bf{k}} + {\bf{q}}) )  
\end{equation}
\begin{equation}
S_{AB}({\bf{k}},t ; {\bf{k}}^{'},t^{'}) 
 = \frac{ |U_{dis}( {\bf{k}} - {\bf{k}}^{'} )|^2 }{ V }
 \frac{ 1 }{ ( \epsilon_{ {\bf{k}} } - \epsilon_{ {\bf{k}}^{'} } )^2  }
\mbox{     }
e^{ -i ( \epsilon_{ {\bf{k}} } - \epsilon_{ {\bf{k}}^{'} } )(t-t^{'}) }
n_{F}({\bf{k}}^{'})(1-n_{F}({\bf{k}}))  
\end{equation}
\begin{equation}
S_{BA}({\bf{k}},t ; {\bf{k}}^{'},t^{'}) 
 = \frac{ |U_{dis}( {\bf{k}}^{'} - {\bf{k}} )|^2 }{ V }
 \frac{ 1 }{ ( \epsilon_{ {\bf{k}} } - \epsilon_{ {\bf{k}}^{'} } )^2  }
\mbox{     }
e^{ -i ( \epsilon_{ {\bf{k}}^{'} } - \epsilon_{ {\bf{k}} } )(t-t^{'}) }
n_{F}({\bf{k}})(1-n_{F}({\bf{k}}^{'}))  
\end{equation}
 In an earlier preprint we showed that the real part of the
 a.c. conductivity may be written as,
\begin{equation}
Re \left[ \sigma(\omega;U_{dis}) \right]
  = \left( \frac{ \pi e^2 }{ m^2 V } \right)
\frac{1}{ \omega } \sum_{ {\bf{k}} {\bf{k}}^{'} }
\sum_{i,j} ({\bf{k}}.{\bf{k}}^{'}) \mbox{     }
{\tilde{N}}({\bf{k}}, \epsilon_{i}, \epsilon_{j};{\bf{k}}^{'},0)
 \mbox{        }\delta(\omega - \epsilon_{i} + \epsilon_{j})
\end{equation}
where,
\begin{equation}
N({\bf{k}},t;{\bf{k}}^{'},0) \equiv
 \left< n_{ {\bf{k}} }(t) n_{ {\bf{k}}^{'} }(0) \right>
 -  \left< n_{ {\bf{k}} }(t) \right> \left< n_{ {\bf{k}}^{'} }(0) \right>
= \sum_{i,j} e^{-i(\epsilon_{i}-\epsilon_{j})t }
{\tilde{N}}({\bf{k}}, \epsilon_{i}, \epsilon_{j};{\bf{k}}^{'},0)
\end{equation}

\begin{equation}
S_{AA}({\bf{k}},t;{\bf{k}}^{'},t^{'}) = 
\delta_{ {\bf{k}}, {\bf{k}}^{'} }
\sum_{i,j} \delta_{ {\bf{k}}_{i}, {\bf{k}} }
\frac{ |U_{dis}({\bf{k}}_{i} - {\bf{k}}_{j})|^2 }{V}
\frac{1}{ (\epsilon_{i} - \epsilon_{j})^2 }
e^{ -i (\epsilon_{i} - \epsilon_{j})(t-t^{'}) }
n_{F}({\bf{k}}_{j}) (1 - n_{F}({\bf{k}}_{i}))
\end{equation}

\begin{equation}
S_{AB}({\bf{k}},t;{\bf{k}}^{'},t^{'}) = 
\sum_{i,j} \delta_{ {\bf{k}}_{j}, {\bf{k}}^{'} }
\delta_{ {\bf{k}}_{i}, {\bf{k}} }
\frac{ |U_{dis}({\bf{k}}_{i} - {\bf{k}}_{j})|^2 }{V}
\frac{1}{ (\epsilon_{i} - \epsilon_{j})^2 }
e^{ -i (\epsilon_{i} - \epsilon_{j})(t-t^{'}) }
n_{F}({\bf{k}}_{j}) (1 - n_{F}({\bf{k}}_{i}))
\end{equation}

\begin{equation}
S_{BA}({\bf{k}},t;{\bf{k}}^{'},t^{'}) = 
\sum_{i,j} \delta_{ {\bf{k}}_{i}, {\bf{k}}^{'} }
\delta_{ {\bf{k}}_{j}, {\bf{k}} }
\frac{ |U_{dis}({\bf{k}}_{i} - {\bf{k}}_{j})|^2 }{V}
\frac{1}{ (\epsilon_{i} - \epsilon_{j})^2 }
e^{ -i (\epsilon_{i} - \epsilon_{j})(t-t^{'}) }
n_{F}({\bf{k}}_{j}) (1 - n_{F}({\bf{k}}_{i}))
\end{equation}

\begin{equation}
S_{BB}({\bf{k}},t;{\bf{k}}^{'},t^{'}) = 
\delta_{ {\bf{k}}, {\bf{k}}^{'} }
\sum_{i,j} \delta_{ {\bf{k}}_{j}, {\bf{k}} }
\frac{ |U_{dis}({\bf{k}}_{i} - {\bf{k}}_{j})|^2 }{V}
\frac{1}{ (\epsilon_{i} - \epsilon_{j})^2 }
e^{ -i (\epsilon_{i} - \epsilon_{j})(t-t^{'}) }
n_{F}({\bf{k}}_{j}) (1 - n_{F}({\bf{k}}_{i}))
\end{equation}

\[
{\tilde{N}}({\bf{k}}, \epsilon_{i}, \epsilon_{j}; {\bf{k}}^{'}, 0)
 = e^{ -4 S^{0}_{A}({\bf{k}}) }
\delta_{ {\bf{k}}, {\bf{k}}^{'} }
\delta_{ {\bf{k}}_{i}, {\bf{k}} }
\frac{ |U_{dis}({\bf{k}}_{i}-{\bf{k}}_{j})|^2 }{V}
\frac{1}{ ( \epsilon_{i} - \epsilon_{j})^2 }
n_{F}({\bf{k}}_{j}) ( 1 - n_{F}({\bf{k}}_{i}) )
\]
\[
 - e^{ -2S^{0}_{A}({\bf{k}}) }e^{ -2 S^{0}_{B}({\bf{k}}^{'}) }
\delta_{ {\bf{k}}_{j}, {\bf{k}}^{'} } \delta_{ {\bf{k}}_{i}, {\bf{k}} }
\frac{ |U_{dis}({\bf{k}}_{i}-{\bf{k}}_{j})|^2 }{V} 
\frac{1}{ (\epsilon_{i} - \epsilon_{j})^2 }
n_{F}({\bf{k}}_{j}) ( 1 - n_{F}({\bf{k}}_{i}) )
\]
\[
 - e^{ -2S^{0}_{B}({\bf{k}}) }e^{ -2 S^{0}_{A}({\bf{k}}^{'}) }
\delta_{ {\bf{k}}_{i}, {\bf{k}}^{'} } \delta_{ {\bf{k}}_{j}, {\bf{k}} }
\frac{ |U_{dis}({\bf{k}}_{i}-{\bf{k}}_{j})|^2 }{V} 
\frac{1}{ (\epsilon_{i} - \epsilon_{j})^2 }
n_{F}({\bf{k}}_{j}) ( 1 - n_{F}({\bf{k}}_{i}) )
\]
\begin{equation}
 + e^{ -4S^{0}_{B}({\bf{k}}) }
\delta_{ {\bf{k}}, {\bf{k}}^{'} } \delta_{ {\bf{k}}_{j}, {\bf{k}} }
\frac{ |U_{dis}({\bf{k}}_{i}-{\bf{k}}_{j})|^2 }{V} 
\frac{1}{ (\epsilon_{i} - \epsilon_{j})^2 }
n_{F}({\bf{k}}_{j}) ( 1 - n_{F}({\bf{k}}_{i}) )
\end{equation}

\section{A.C. Conductivity}

 The disorder averaged
 a.c. conductivity for Gaussian disorder may be writtten as
\[
\sigma(\omega) \equiv < Re \left[ \sigma(\omega;U_{dis}) \right] >_{dis}
\]
\begin{equation}
  = \left( \frac{ \pi e^2 }{ m^2 V } \right)
\frac{1}{ \omega } 
\sum_{i,j} \left( {\bf{k}}^2_{i} \mbox{     }
e^{ -4 S^{0}_{A}({\bf{k}}_{i}) }
 + {\bf{k}}^2_{j} \mbox{     }
e^{ -4 S^{0}_{B}({\bf{k}}_{j}) } \right)
\frac{ \Delta^2 }{V}
\frac{1}{ ( \frac{ {\bf{k}}_{i}^2 }{2m} - \frac{ {\bf{k}}_{j}^2 }{2m })^2 }
n_{F}({\bf{k}}_{j}) ( 1 - n_{F}({\bf{k}}_{i}) )
 \mbox{        }\delta(\omega - \frac{ {\bf{k}}_{i}^2 }{2m}
 + \frac{ {\bf{k}}_{j}^2 }{2m} )
\end{equation}
where,
\begin{equation}
S^{0}_{A}({\bf{k}}) = \sum_{ {\bf{q}} } \frac{ \Delta^2 }{V}
\frac{ m^2 }{ ( {\bf{k.q}} - {\bf{q}}^2/2 )^2 }
n_{F}({\bf{k}}-{\bf{q}}) (1- n_{F}({\bf{k}}))
\end{equation}
\begin{equation}
S^{0}_{B}({\bf{k}}) = \sum_{ {\bf{q}} } \frac{ \Delta^2 }{V}
\frac{ m^2 }{ ( {\bf{k.q}} + {\bf{q}}^2/2 )^2 }
n_{F}({\bf{k}}) (1- n_{F}({\bf{k}}+{\bf{q}}))
\end{equation}
In other words,
\[
\sigma(\omega) \equiv < Re \left[ \sigma(\omega;U_{dis}) \right] >_{dis}
\]
\begin{equation}
  = \left( \frac{ 2 \pi e^2 \Delta^2 }{ m } \right)
\frac{1}{ \omega^3 } 
 \int^{ \epsilon_{F} }_{ \epsilon_{F} - \omega }
 d \epsilon \mbox{     }
D(\omega+\epsilon)
D(\epsilon)  \mbox{     }
 \left( (\omega + \epsilon) \mbox{     }
e^{ -4 S^{0}_{A}(\omega+\epsilon) }
 + \epsilon \mbox{     }
e^{ -4 S^{0}_{B}(\epsilon) } \right) 
\end{equation}
\begin{equation}
S^{0}_{A}(\epsilon) =  \Delta^2 \int^{ \epsilon_{F} }_{0} d \epsilon^{'} 
 D(\epsilon^{'}) \mbox{         }  
\frac{1}{ ( \epsilon - \epsilon^{'} )^2 }
\end{equation}
\begin{equation}
S^{0}_{B}(\epsilon) =  \Delta^2 \int^{\infty}_{ \epsilon_{F} } d \epsilon^{'} 
 D(\epsilon^{'}) \mbox{         }  
\frac{1}{ (\epsilon^{'} - \epsilon)^2 }
\end{equation}
\begin{equation} 
D(\epsilon) = \frac{m}{ (2 \pi)^d } (2m\mbox{     } \epsilon)
^{ \frac{ d-2 }{2} } 
\end{equation}
Using $ Mathematica^{TM} $ we find,
\begin{equation}
S^{0}_{A}(\epsilon) = \Delta^2 \frac{ m }{ (2 \pi)^d }
(2m)^{ \frac{ d-2 }{2} }
\left( \frac{ 2 \epsilon_{F}^{d/2} }{d \mbox{      }\epsilon^2 } \right)
\mbox{        }H2F1[ 2,  d/2,  1 + d/2, \frac{ \epsilon_{F} }{ \epsilon } ]
\end{equation}
\begin{equation}
S^{0}_{B}(\epsilon) = \Delta^2 \frac{ m }{ (2 \pi)^d }
(2m)^{ \frac{ d-2 }{2} }
\left( \frac{ -2 \epsilon_{F}^{ (-4+d)/2 } }
{ (-4 + d) } \right)
\mbox{        }
H2F1[ 2,  2 - d/2,  3 - d/2, \frac{ \epsilon }{ \epsilon_{F} } ]
\end{equation}
In one dimension, further simplification is not possible.
Hence we write,
\begin{equation}
S^{0}_{A}(\epsilon) = \Delta^2 \frac{ m }{ (2 \pi) }
(2m)^{ -\frac{ 1 }{2} }
\left( \frac{ 2 \epsilon_{F}^{1/2} }{ \epsilon^2 } \right)
\mbox{        }H2F1[ 2,  1/2,  3/2, \frac{ \epsilon_{F} }{ \epsilon } ]
\end{equation}
\begin{equation}
S^{0}_{B}(\epsilon) = \Delta^2 \frac{ m }{ (2 \pi) }
(2m)^{ -\frac{ 1 }{2} }
\left( \frac{ 2 \epsilon_{F}^{ -3/2 } }
{ 3 } \right)
\mbox{        }
H2F1[ 2,  3/2,  5/2, \frac{ \epsilon }{ \epsilon_{F} } ]
\end{equation}
In two spatial dimensions we have,
\begin{equation}
S^{0}_{A}(\epsilon) = \Delta^2 
\frac{ m }{ (2 \pi)^2 } \left( \frac{1}{ \epsilon - \epsilon_{F} }
 - \frac{1}{ \epsilon }  \right)
\end{equation}
\begin{equation}
S^{0}_{B}(\epsilon) = \Delta^2 \frac{ m }{ (2 \pi)^2 }
\mbox{        }
\frac{ 1 }{ \epsilon_{F} - \epsilon }
\end{equation}
In three spatial dimensions we have,
\begin{equation}
S^{0}_{A}(\epsilon) = \Delta^2 \frac{ m }{ (2 \pi)^3 }
(2m)^{ \frac{ 1 }{2} }
\left( \frac{ 2 \epsilon_{F}^{3/2} }{3 \mbox{      }\epsilon^2 } \right)
\mbox{        }H2F1[ 2,  3/2,  5/2, \frac{ \epsilon_{F} }{ \epsilon } ]
\end{equation}
\begin{equation}
S^{0}_{B}(\epsilon) = \Delta^2 \frac{ m }{ (2 \pi)^3 }
(2m)^{ \frac{ 1 }{2} }
\left( 2 \epsilon_{F}^{ -1/2 }  \right)
\mbox{        }
H2F1[ 2,  1/2,  3/2, \frac{ \epsilon }{ \epsilon_{F} } ]
\end{equation}
In two dimensions we have,
\begin{equation}
\sigma(\omega) 
  = \left( \frac{ 2 \pi e^2 \Delta^2 }{ m } \right)
\frac{1}{ \omega^3 } \frac{ m^2 }{ (2 \pi)^4 } 
 \int^{ \epsilon_{F} }_{ \epsilon_{F} - \omega }
 d \epsilon \mbox{     }
 \left( (\omega + \epsilon) \mbox{     }
e^{ -4  \Delta^2
\frac{ m }{ (2 \pi)^2 } \left( \frac{1}{ \epsilon + \omega - \epsilon_{F} }
 - \frac{1}{ \epsilon + \omega }  \right)  }
 + \epsilon \mbox{     }
e^{ -4 \Delta^2  \frac{ m }{ (2 \pi)^2 }
\mbox{        }
\frac{ 1 }{ \epsilon_{F} - \epsilon }  } \right) 
\end{equation}
 This may be approximately evaluated as follows.
\begin{equation}
\sigma(\omega) 
  \approx \left( \frac{ 2 \pi e^2 \Delta^2 }{ m } \right)
\frac{1}{ \omega^2 } \frac{ m^2 }{ (2 \pi)^4 } 
 \left( 2\epsilon_{F} \mbox{     }
e^{ -4  \Delta^2
\frac{ m }{ (2 \pi)^2 } \frac{2}{ \omega }  } \right) 
\end{equation}
 It can bee seen that
 the zero frequency limit of the above expression is zero since the 
 integral vanishes exponentially fast $ \sim e^{-c_{0}/\omega }/\omega^2 $.
 Thus
 the d.c. conductivity of a two dimensional system is zero and the frequency
 dependence is rather nontrivial. Similarly we may expect that
 in one dimension the d.c. conductivity vanishes.  Unfortunately 
 for a similar reason we find that the d.c. conductivity in three
 dimensions also vanishes. This means we have to include
 terms beyond what Eq.(~\ref{TOYHAMIL}) does. Perhaps the reader can do
 this or at least offer to collaborate with the author. Please contact
 me at gsetlur@imsc.res.in

\section{Some Technical Musings}

 It appears that the mathematical literature on the subject of 
 quantum particles in random potentials is vast\cite{Krishna}.  
 It is possible, indeed likely that many mathematically rigorous 
 results are known regarding this problem. But this does not prevent
 the authors from making some remarks that more knowledgeable
 readers may choose to critique. In particular, the author is
 uncomfortable with the notion of disorder averaging. Nature
 chooses its potentials based on the distribution of impurities, defects and
 so on. This potential is fixed and well-defined for a particular distribution
 of these imperfections. The physicists' ignorance of the precise nature
 of this potential is not a license to average over these potentials. 
 Nature does not average, people do. But are people justified in averaging ?
 In other words can averaging simplify the problem without washing out
 essential physics ? In order to answer this question we have to make
 the following conjectures. 

\vspace{0.1in}

\noindent {\bf{Defn0}} : Let $ {\mathcal{U}}_{d} $ be the set of
 all potentials $ U({\bf{x}}) $ in a fixed spatial dimension $ d $.

\noindent {\bf{Defn1}} : Let $ {\mathcal{F}}_{d} $ be the set of
 all potentials $ U({\bf{x}}) $ in a fixed spatial dimension $ d $
 that has the following property. They all lead to the same
 exponent $ \delta $ for the frequency dependence of the a.c. conductivity.
 In other words, each of these potentials predicits that
 $ Re[ \sigma(\omega) ] \sim \omega^{\delta} $
 (in some region of $ \omega $ with possibly
 some additive part independent of $ \omega $)
  with the {\it{same}} $ \delta $.

\noindent {\bf{Conjecture 1}} :  $ {\mathcal{F}}_{d} $ is dense in
 $ {\mathcal{U}}_{d} $. 

 If $ {\bf{Conjecture 1}} $ is valid, then one may average over all these
 `sufficiently erratic' potentials and expect to extract $ \delta $ which
 is all that physicists care about. It is possible that $ \delta $ may
 be extracted from a numerical solution of the Schrodinger equation 
  using a specific $ U $ that 
 belongs to the set $  {\mathcal{F}}_{d} $. But this would involve
 using the computer for more than checking one's email, 
 and not everyone likes that.

\noindent {\bf{Defn2}} : Let $ {\mathcal{M}}_{3} $ be the set of
 all potentials $ U({\bf{x}}) $ in spatial dimension $ d = 3 $
 that has the following property. They all lead to the same
 exponent $ \beta $ for the mobility edge exponent.
 In other words, each of these potentials predicit that
 $ \sigma_{d.c.} \sim (E_{F} - E_{c})^{\beta}
 \mbox{     }\theta(E_{F}-E_{c}) $ with the {\it{same}} $ \beta $.
 However for different potentials, $ E_{c} $ - the mobility
 edge, may be different.

\noindent {\bf{Conjecture 2}} :  $ {\mathcal{F}}_{3} $ is dense in
 $ {\mathcal{U}}_{3} $. 

 If $ {\bf{Conjecture 2}} $ is valid, then one may average over all these
 `sufficiently erratic' potentials and expect to extract $ \beta $.

 Thus the validity of the process of averaging over potentials rests crucially
 it seems, on all these sufficiently erratic potentials predicting the same
 exponents and on these sufficiently erratic potentials spanning nearly all
 possible potentials. 

 If both these are satistifed then one may average over all potentials and
 extract the exponents, or, if one is better at programming, 
 choose a particular potential from this set, numerically solve the
 Schrodinger equation and extract the exponent from there. In either
 case we should get the same answer. A final conjecture seems appropriate.

 {\bf{ Conjecture 3 }} : Let $ {\mathcal{M}}^{'}_{3} $ have an exponent
 $ \beta^{'} $ and  $ {\mathcal{F}}^{'}_{d} $ have an exponent $ \delta^{'} $,
 then $ \beta = \beta^{'} $ and $ \delta = \delta^{'} $. In other words,
 these exponents are unique.

 With powerful computers now available, purely analytical methods such as this
 work may seem  pass\`{e},
 but a closed formula for the a.c. conductivity
 that one can stare at (and one that is hopefully right) and admire
 has a charm that a cold data file on the hard disk is unable to 
 duplicate. Besides, with Coulomb interaction, the problem becomes intractable
 numerically, however, one may expect to combine the sea-boson method with
 the present one to extract the exponents analytically.

\end{document}